\documentstyle[12pt]{article}

\newtheorem{theorem}{Theorem}
\newenvironment{acknowledge}%
  {\bigskip\noindent{\bf Acknowledgements}\bigskip}%
  {\bigskip}
\newcommand{\beqn}{\begin{equation}}
\newcommand{\eeqn}{\end{equation}}
\newcommand{\beqnarray}{\begin{eqnarray}}
\newcommand{\eeqnarray}{\end{eqnarray}}
\newcommand{\rd}{\partial}
\newcommand{\dfrac}[2]{ \frac{\displaystyle #1}{\displaystyle #2} }

\newcommand{\res}{\mathop{\mbox{\rm res}}}

\newcommand{\cB}{{\cal B}}
\newcommand{\cBbar}{{\bar{\cB}}}

\newcommand{\Edot}{{\dot{E}}}
\newcommand{\cF}{{\cal F}}
\newcommand{\cL}{{\cal L}}
\newcommand{\cLbar}{{\bar{\cL}}}
\newcommand{\cM}{{\cal M}}
\newcommand{\cMbar}{{\bar{\cM}}}

\newcommand{\Pbar}{{\bar{P}}}

\newcommand{\qbar}{{\bar{q}}}
\newcommand{\Sbar}{{\bar{S}}}
\newcommand{\cT}{{\cal T}}
\newcommand{\tbar}{{\bar{t}}}

\newcommand{\vbar}{{\bar{v}}}
\newcommand{\lambdabar}{{\bar{\lambda}}}
\newcommand{\mubar}{{\bar{\mu}}}
\newcommand{\xibar}{{\bar{\xi}}}
\begin{document}
\date{ }
\title{Dispersionless Hierarchies, Hamilton-Jacobi Theory 
and Twistor Correspondences}
\author{
    Partha Guha$^1$
    \footnote{On leave from S.N. Bose National Center 
    for Basic Science, Salt-Lake, Culcutta-91, India}
    and Kanehisa Takasaki$^2$\\
\\
$^1${\normalsize 
Research Institute for Mathematical Sciences, Kyoto University}\\
{\normalsize Sakyo-ku, Kyoto 606, Japan}\\
$^2${\normalsize 
Department of Fundamental Sciences, Kyoto University}\\
{\normalsize Sakyo-ku, Kyoto 606, Japan}
}
\maketitle
\bigskip

\begin{center}
    {\it Dedicated to Sir Roger Penrose on his 65th birthday}
\end{center}
\bigskip

\begin{abstract}
\noindent The dispersionless KP and Toda hierarchies possess 
an underlying twistorial structure. A twistorial approach is 
partly implemented by the method of Riemann-Hilbert problem. 
This is however still short of clarifying geometric ingredients 
of twistor theory, such as twistor lines and twistor surfaces. 
A more geometric approach can be developed in a Hamilton-Jacobi 
formalism of Gibbons and Kodama. \\
{\bf AMS Subject Classifiation} (1991): 35Q20, 58F07,70H99
\end{abstract}

\newpage

\section{Introduction}

The dispersionless KP and Toda hierarchies are the 
most typical dispersionless integrable hierarchies. 
The ordinary KP and Toda hierarchies have a Lax 
representation of the form 
\beqn
    \dfrac{\rd L}{\rd t_n} = [B_n, L], \quad 
    \mbox{etc} \ldots
\eeqn
where $L,B_n$, etc. are pseudo-differential or difference 
operators. Their dispersionless limit is a kind of 
``quasi-classical'' limit in which $L,B_n$, etc. are 
phase space functions and the commutator $[\ ,\ ]$ is 
replaced by a Poisson bracket $\{\ ,\ \}$. This causes 
a drastic change, especially in the Toda hierarchy -- 
the one-dimensional lattice in the Toda hierarchy turns 
into a continuous line.  The dispersionless hierarchies 
nevertheless inherit many aspects of integrability from 
the original KP and Toda hierarchies, such as the notion 
of $\tau$ functions, Hirota equations, infinite-dimensional 
symmetries, etc.  (For an overview, we refer to the review
\cite{takasaki-takebe-95}.)  

In some aspects, however, the dispersionless hierarchies 
require an entirely new approach.  One of them is the problem 
of constructing special solutions.  A standard recipe is to 
consider a suitable ``reduction'' of the original infinite 
hierarchy into a system with a finite number of unknown 
functions. Usually, such a reduced system of the dispersionless 
hierarchies is a hydrodynamic system, and solvable by a 
generalized ``hodographic'' method as developed by Tsarev 
\cite{tsarev-85}.  Many special solutions of the dispersionless 
KP and Toda hierarchies have been indeed constructed by 
Gibbons and Kodama \cite{kodama-88,gibbons-kodama-89,kodama-90} 
by a generalized hodographic method. 

Proof of integrability of the dispersionless hierarchies 
themselves, too, has to be established by a new method. 
Such an approach is provided by the method of Riemann-Hilbert 
problem \cite{takasaki-takebe-91,takasaki-takebe-92a}. This 
Riemann-Hilbert problem is discovered as an analogue of the 
Riemann-Hilbert problem for the four-dimensional self-dual 
Einstein equation \cite{boyer-plebanski-85} and an associated 
infinite hierarchy \cite{takasaki-90}. In the case of the 
self-dual Einstein equation, the Riemann-Hilbert problem is 
an analytic expression of Penrose's curved twistor theory 
\cite{penrose-76}. Thus the method of Riemann-Hilbert method 
may be called a twistorial approach to the dispersionless 
hierarchies. (For more details on this analogy, we refer to 
the review \cite{takasaki-92}.) 

Although thus nicely exhibiting a link with twistor theory, 
however, the previous approach by the Riemann-Hilbert problem 
\cite{takasaki-takebe-91,takasaki-takebe-92a} 
is still lacking a geometric language.  Let us recall that 
a central idea of twistor theory is to relate a space-time 
manifold $M$ with a complex manifold $\cT$ (twistor space) by 
the ``twistor correspondence'', i.e., a manifold $\cF$ with a 
double fibration $\pi_1: \cF \to M$ and $\pi_2: \cF \to \cT$. 
By this correspondence, a space-time point $x \in M$ 
determines (and is determined by) a rational curve (twistor 
line) $\pi_2(\pi_1^{-1}(x))$ in the twistor space, and a 
twistor point $\xi \in \cT$ a submanifold (twistor surface)
$\pi_1(\pi_2^{-1}(\xi))$ of the space-time.  These notions 
have been left obscure in the case of the dispersionless 
hierarchies. 

The goal of this paper is to establish a dictionary between 
the twistor geometry and the method of Riemann-Hilbert problem 
for both the dispersionless KP and Toda hierarchy.  It turns 
out that the Hamilton-Jacobi approach of Gibbons and Kodama 
\cite{kodama-gibbons-90,gibbons-kodama-94}, another proof of 
integrability of the dispersionless KP hierarchy, provides 
such a geometric interpretation of the method of Riemann-Hilbert 
problem. In Section 2, we review the previous results on the 
dispersionless KP hierarchy, and point out the problem. 
In Section 3, we reformulate the Hamilton-Jacobi approach 
of Gibbons and Kodama in our language, and present the 
twistorial interpretation. Section 4 is devoted to a 
similar treatment of the dispersionless Toda hierarchy. 
This part may be read as an independent result on an 
extension of the approach of Gibbons and Kodama. 
Section 5 is added for comments on a relation to Frobenius 
structures.

\section{Dispersionless KP hierarchy and Twistor Theory}

\subsection{Lax formalism} 

Let us recall the Lax formalism of the dispersionless KP hierarchy 
\cite{krichever-92,takasaki-takebe-92a}. 

Let $(p,x)$ be canonical coordinates in a two dimensional 
phase space with Poisson bracket 
\beqn
    \{ F, G \} = \frac{\rd F}{\rd p} \frac{\rd G}{\rd x}
               - \frac{\rd F}{\rd x} \frac{\rd G}{\rd p}. 
\eeqn
Let $t = (t_2,t_3,\ldots)$ be a set of ``time variables''. 
The dispersionless KP hierarchy has the Lax representation 
\beqn
    \dfrac{\rd \cL}{\rd t_n} = \{ \cB_n, \cL\}, 
\eeqn
where $\cL$ is a Laurent series of $p$ of the form 
\beqn
    \cL = p + \sum_{n=1}^\infty u_{n+1}(x,t) p^{-n}. 
\eeqn
$\cB_n$ are the polynomial (in $p$) part of the $n$-th 
power of $\cL$, 
\beqn
    \cB_n = \Bigl( \cL^n \Bigr)_{\ge 0}, 
\eeqn
and obey the Zakharov-Shabat equations 
\beqn
    \dfrac{\rd \cB_m}{\rd t_n} - \dfrac{\rd \cB_n}{\rd t_m} 
    + \{ \cB_m, \cB_n \} = 0. 
\eeqn

One can extend these Lax equations by adding another 
Laurent series $\cM$ of the form 
\beqn
    \cM = \sum_{n=2}^\infty n t_n \cL^{n-1} + x 
            + \sum_{n=1}^\infty v_n(x,t) \cL^{-n-1} 
\eeqn
that satisfies the Lax equations 
\beqn
    \dfrac{\rd \cM}{\rd t_n} = \{ \cB_n, \cM\} 
\eeqn
and the canonical Poisson bracket relation 
\beqn
    \{ \cL, \cM \} = 1. 
\eeqn

The dispersionless KP hierarchy is thus a ``quasi-classical''
limit of the ordinary KP hierarchy with commutators replaced by 
Poisson brackets. $\cL$ corresponds to the Lax operator of the 
ordinary Lax formalism of the KP hierarchy. $\cM$ is the 
dispersionless version of the Orlov-Schulman operator $M$ 
\cite{orlov-schulman-86}.

\subsection{Twistorial structure}

The extended Lax formalism has another expression 
\cite{takasaki-takebe-92a} which resembles twistor theory. 
Let $\omega$ be the 2-form
\beqn
    \omega = dp \wedge dx + \sum_{n=2}^\infty d\cB_n \wedge dt_n. 
\eeqn
In terms of this 2-form, the Zakharov-Shabat equations and 
the Lax equations can be rewritten in a very compact form as 
\beqn
    \omega \wedge \omega = 0, 
\eeqn
and 
\beqn
    d\cL \wedge d\cM = \omega, 
\eeqn
respectively. 
These equations show that $\omega$ is a degenerate symplectic 
form (on the infinite dimensional $(p,x,t)$ space) and that 
$\cL$ and $\cM$ are its Darboux variables. The 2-form $\omega$ 
is an analogue of Gindikin's ``bundle of 2-form'', which lies 
in the heart of his reformulation of curved twistor theory
\cite{gindikin-82}.  This suggests that the dispersionless 
KP hierarchy will give a kind of curved twistor theory. 

This analogy is further deepend by the following characterization 
of $\cL$ and $\cM$ in terms of a Riemann-Hilbert problem 
\cite{takasaki-takebe-92a}: 

\begin{theorem}
The functions $\cL = \cL(p,x,t)$ and $\cM = \cM(p,x,t)$ 
obeys the functional equations (Riemann-Hilbert problem) 
\beqnarray
    \cL(p,x,t) &=& f\Bigl( \cLbar(p,x,t), \cMbar(p,x,t) \Bigr), 
                                                       \nonumber \\
    \cM(p,x,t) &=& g\Bigl( \cLbar(p,x,t), \cMbar(p,x,t) \Bigr), 
\eeqnarray
where $\cLbar = \cLbar(p,x,t)$ and $\cMbar = \cMbar(p,x,t)$ are 
a (unique) solution of the equations 
\beqnarray
  &&
    \dfrac{\rd \cLbar}{\rd t_n} = \{ \cB_n, \cLbar \}, \quad 
    \dfrac{\rd \cMbar}{\rd t_n} = \{ \cB_n, \cMbar \}, 
    \nonumber \\
  &&
    \{ \cLbar, \cMbar \} = 1, 
\eeqnarray
under the initial condition
\beqn
    \cLbar(p,x,0) = p, \quad \cMbar(p,x,0) = x, 
\eeqn
and $f = f(p,x)$ and $g = g(p,x)$ are given by 
\beqn
    f(p,x) = \cL(p,x,0), \quad g(p,x) = \cM(p,x,0). 
\eeqn
\end{theorem}

{\bf Proof}: Note that $f$ and $g$  give a canonical pair 
\beqn
    \{ f, g \} = 1.  
\eeqn Therefore $\cL$ and $f(\cLbar,\cMbar)$ turn out to satisfy the same
Lax equations. Furthermore they have an identical initial value at $t = 0$. 
Thus they coincide. Similarly, we find that $\cM = g(\cLbar,\cMbar)$. 
$\Box$ 
\bigskip

Let us specify why the above functional equations are interpreted 
as Riemann-Hilbert problem. In a suitable analytical setting, 
$\cLbar$ and $\cMbar$ are holomorphic functions of $p$ in a 
neighborhood $\bar{D}$ of $p = 0$ (because $\cB_n$ are polynomials 
of $p$), whereas $\cL$ and $\Xi = \cM - \sum n t_n \cL^{n-1}$ 
are holomorphic functions in a neighborhood $D$ of $p = \infty$. 
Suppose that $D$ and $\bar{D}$ cover of the whole Riemann sphere 
(e.g., this is the case if $t$ is small).  A precise meaning of 
the above functional equations is that these four functions 
$(\cL,\Xi,\cLbar,\cMbar)$ satisfy the equation 
\beqn 
    \cL = f(\cLbar,\cMbar), \quad 
    \Xi = g(\cLbar,\cMbar) 
      - \sum_{n=2}^\infty n t_n f(\cLbar,\cMbar)^{n-1} 
\eeqn
over the intersection $D \cap \bar{D}$. This is a Riemann-Hilbert 
problem of the type that typically arise in curved twistor 
theory \cite{penrose-76,boyer-plebanski-85}. 

If the flows are restricted to the $N$-dimensional subspace 
$(x,t) = (x,t_1,\ldots,t_N,0,\ldots)$, the analogy with 
curved twistor theory becomes more rigorous. One can indeed 
``twist'' $(f,g)$ into a map connecting $(\cL^N, \cM \cL^{1-N}/N)$ 
and $(\cLbar,\cMbar)$, use it as a transition function for 
constructing a mini-twistor space \cite{hitchin-82,jones-tod-85} 
by gluing two coordinate patches. The four functions then become 
``twistor functions''.  

In the full space of time evolutions (i.e., if $N = \infty$), 
the mini-twistor space ceases to exist. The Riemann-Hilbert 
problem itself is meaningful, but geometric ingredients 
of twistor theory,  such as ``twistor correspondence'', 
``twistor lines'' and ``twistor surfaces'' remain to be 
clarified.  This is the problem that we are addressing. 

Our claim is that the Hamilton-Jacobi approach of Gibbons 
and Kodama \cite{kodama-gibbons-90,gibbons-kodama-94} may be used 
to resolve this problem.

\section{Hamilton-Jacobi Theory and Twistor Geometry}

\subsection{Multi-time Hamiltonian system}

In this and next subsections, we reformulate the Hamilton-Jacobi 
approach of Gibbons and Kodama in our language. See also Carroll's 
paper \cite{carroll-94} which presents an exposition of the 
Hamilton-Jacobi approach in a form more faithful to the formulation 
of Gibbons and Kodama. 

Let $\lambda$ and $\mu$ be parameters, and define two functions 
$p = p(\lambda,\mu,t)$ and $x = x(\lambda,\mu,t)$ implicitly 
by the equations 
\beqn
    \cL(p,x,t) = \lambda, \quad 
    \cM(p,x,t) = \mu.
\eeqn

\begin{theorem}
$p = p(\lambda,\mu,t)$ and $x = x(\lambda,\mu,t)$ satisfy 
the multi-time Hamiltonian system 
\beqn
    \dfrac{dp}{dt_n} = \dfrac{\rd \cB_n}{\rd x}, \quad 
    \dfrac{dx}{dt_n} = - \dfrac{\rd \cB_n}{\rd p} 
\eeqn
with time-dependent Hamiltonians $\cB_n = \cB_n(p,x,t)$. 
\end{theorem}

{\bf Proof}:  For avoiding complicated notations, let us write 
$p(\lambda,\mu,t)$ and $x(\lambda,\mu,t)$ as 
$p(t)$ and $x(t)$. By the definition, we have the identities
\beqn
    \cL\Bigl( p(t), x(t), t \Bigr) = \lambda, \quad 
    \cM\Bigl( p(t), x(t), t) \Bigr) = \mu.          \nonumber 
\eeqn
Differentiating these identities in $t_n$ gives 
\beqnarray
    \dfrac{\rd \cL}{\rd p} \dfrac{dp}{dt_n} + 
    \dfrac{\rd \cL}{\rd x} \dfrac{dx}{dt_n} + 
    \dfrac{\rd \cL}{\rd t_n}                   = 0,  
                                               \nonumber \\
    \dfrac{\rd \cM}{\rd p} \dfrac{dp}{dt_n} + 
    \dfrac{\rd \cM}{\rd x} \dfrac{dx}{dt_n} + 
    \dfrac{\rd \cM}{\rd t_n}                   = 0. 
\eeqnarray
The Lax equations for $\cL$ and $\cM$ can be used here to 
evaluate the last terms in these equations.  This gives 
\beqnarray
    \left( \begin{array}{rr}
       \dfrac{\rd \cL}{\rd p} & \dfrac{\rd \cL}{\rd x} \\
       \dfrac{\rd \cM}{\rd p} & \dfrac{\rd \cM}{\rd x} 
    \end{array} \right) 
    \left( \begin{array}{r}
       \dfrac{\rd dp}{\rd t_n} \\
       \dfrac{\rd dx}{\rd t_n}
    \end{array} \right) 
    =  
    \left( \begin{array}{r}
       - \dfrac{\rd \cL}{\rd t_n} \\
       - \dfrac{\rd \cM}{\rd t_n}
    \end{array} \right)              \nonumber \\
    = 
    \left( \begin{array}{rr}
       \dfrac{\rd \cL}{\rd p} & \dfrac{\rd \cL}{\rd x} \\
       \dfrac{\rd \cM}{\rd p} & \dfrac{\rd \cM}{\rd x} 
    \end{array} \right) 
    \left( \begin{array}{r}
       \dfrac{\rd \cB_n}{\rd x} \\
       - \dfrac{\rd \cB_n}{\rd p}
    \end{array} \right). 
\eeqnarray 
The common $2 \times 2$ array is invertible (because of the 
canonical Poisson commutation relation of $\cL$ and $\cM$) 
and can be removed. The resulting equations are what we have 
sought for. 
$\Box$
\bigskip

The multi-time Hamiltonian system of Gibbons and Kodama can be 
thus reproduced from our extended Lax formalism. 

This result can be restated in terms of canonical transformations. 
The Hamiltonian system is derived as equations of motion of 
a point $(p,x)$ keeping $\cL$ and $\cM$ constant. In other words, 
$\cL$ and $\cM$ are invariants of the Hamiltonian flows: 
\beqnarray
    \cL\Bigl( p(t),x(t),t \Bigr) &=& \cL\Bigl( p(0),x(0),0 \Bigr), 
                                                     \nonumber \\
    \cM\Bigl( p(t),x(t),t \Bigr) &=& \cM\Bigl( p(0),x(0),0 \Bigr). 
\eeqnarray
Meanwhile, by the canonical Poisson commutation relation between 
$\cL$ and $\cM$, the two-dimensional map $(p,x) \mapsto 
(\lambda,\mu) = \Bigl(\cL(p,x,t),\cM(p,x,t)\Bigr)$ is symplectic. 
The above invariance property of $\cL$ and $\cM$ then implies that 
this symplectic map is a canonical transformation converting the 
multi-time Hamiltonian system to the Hamiltonian system 
\beqn
    \dfrac{d\lambda}{dt_n} = 0, \quad 
    \dfrac{d\mu}{dt_n} = 0 
\eeqn
with {\it zero Hamiltonians}. This Hamiltonian system is further 
transformed into the Hamiltonian system 
\beqn
    \dfrac{d\lambda}{dt_n} = 0, \quad 
    \dfrac{d\xi}{dt_n} = - n \lambda^{n-1}
\eeqn
by the simple transformation
\beqn
    \xi = \mu - \sum_{n=2}^\infty n t_n \lambda^{n-1}. 
\eeqn
(Recall that the last transformation is just a disguise of the 
transformation $\Xi = \cM - \sum n t_n \cL^{n-1}$.) The last 
canonical variables $(\lambda,\xi)$ are exactly those of Gibbons 
and Kodama. As they pointed out, the Hamiltonian system in these 
variables resembles the time evolution of {\it scattering data} 
in the conventional inverse scattering problem, which are also 
{\it action-angle variables}.

\subsection{Generating function} 

Gibbons and Kodama formulated the above canonical transformation 
$(p,x) \mapsto (\lambda,\xi)$ in terms of a generating function. 
In our formulation, it is more convenient to consider the 
canonical transformation $(p,x) \mapsto (\lambda,\mu)$. Let 
$S(\lambda,x,t)$ be a generating function for the latter. 
The relations among the canonical variables $(p,x)$, 
$(\lambda,\mu)$ and the Hamiltonian $\cB_n$ are now written 
\beqnarray
  && \dfrac{\rd S(\lambda,x,t)}{\rd \lambda} = \mu, \quad 
     \dfrac{\rd S(\lambda,x,t)}{\rd x} = p, \nonumber \\
  && \dfrac{\rd S(\lambda,x,t)}{\rd t_n} = \cB_n. 
\eeqnarray
These relations can be cast into a more compact form: 
\beqn
    dS = \mu d\lambda + pdx + \sum_{n=2}^\infty \cB_n dt_n. 
\eeqn
This shows that the generating function is nothing but the 
$S$-function that was discovered in an entirely different context 
\cite{krichever-92,takasaki-takebe-92a}. 
The generating function of Gibbons and Kodama, say 
$S_{GK}$, is related to ours as 
$S_{GK} = S - \sum \lambda^n t_n$.

The generating function $S(\lambda,x,t)$ is also related to the 
linear problem 
\beqnarray
  && \lambda \Psi = L\Bigl(\hbar\frac{\rd}{\rd x}\Bigr)\Psi, \quad 
     \hbar \frac{\rd \Psi}{\rd \lambda} 
     = M\Bigl(\hbar\frac{\rd}{\rd x}\Bigr)\Psi, \nonumber \\
  && \hbar \dfrac{\rd \Psi}{\rd t_n} 
     = B_n\Bigl(\hbar\frac{\rd}{\rd x}\Bigr)\Psi
\eeqnarray
of the KP hierarchy \cite{takasaki-takebe-92b,takasaki-takebe-95}. 
The WKB approximation
\beqn
    \Psi \sim \exp \hbar^{-1}S(\lambda,x,t). 
\eeqn
to this {\it multi-time Schr\"odinger system} gives a system of 
Hamilton-Jacobi (or ``eikonal'') equations of the form
\beqnarray
  &&
    \lambda = \cL\Bigl(\frac{\rd S(\lambda,x,t)}{\rd x}\Bigr), 
  \quad 
    \dfrac{\rd S(\lambda,x,t)}{\rd \lambda}  
    = \cM\Bigl(\frac{\rd S(\lambda,x,t)}{\rd x}\Bigr) \nonumber \\
  &&
    \dfrac{\rd S(\lambda,x,t)}{\rd t_n} 
    = \cB_n\Bigl(\frac{\rd S(\lambda,x,t)}{\rd x}\Bigr). 
\eeqnarray
These Hamilton-Jacobi equations turn into the above defining 
relations of $S(\lambda,x,t)$ by the change of coordinates 
$(\lambda,x,t) \to (p,x,t)$ with 
\beqn
    p = \frac{\rd S(\lambda,x,t)}{\rd x}. 
\eeqn
Thus, the generating function $S(\lambda,x,t)$ links 
a {\it particle picture} (multi-time Hamiltonian system) and 
a {\it wave picture} (multi-time Schr\"odinger system).

\subsection{Twistor geometry} 

Let us now compare these results with curved twistor 
theory. As the Riemann-Hilbert problem suggests, 
$\lambda$ and $\mu$ may be thought of as coordinates on a 
(virtual) twistor space, and $\cL$ and $\cM$ as defining a 
twistor correspondence between space-time points and twistor 
points. According to the ordinary curved twistor theory 
of 4D space-times \cite{penrose-76}, the level surfaces 
of $\cL$ and $\cM$ should be twistor surfaces.  Actually, 
the present setting rather resembles the curved twistor theory 
of 3D space-times\cite{hitchin-82,jones-tod-85}. From that 
point of view, it seems more suitable to interpret $(x,t)$ 
as space-time coordinates and $p$ as the fiber coordinate 
of a projectivized ``spinor bundle''.  Accordingly, $p(t)$ 
is a ``covariantly constant spinor field'' evaluated at 
the space-time point $(x,t) = (x(t),t)$ of a twistor 
surface. 

Summarizing, we have the following dictionary between 
the dispersionless KP hierarchy and the twistor geometry 
(in particular, the space-time side): 
\begin{itemize}
\item $(\lambda,\mu)$ $\longleftrightarrow$ mini-twistor space
\item $(x,t)$ $\longleftrightarrow$ space-time
\item $p$ $\longleftrightarrow$ fiber of projectivized spinor bundle
\item $(x(t),t)$ $\longleftrightarrow$ twistor surface
\item $p(t)$ $\longleftrightarrow$ covariantly constant spinor 
field on this twistor surface 
\end{itemize}
Of course the notion of mini-twistor space is only virtual, but 
the others have a definite meaning. 

Having obtained this twistorial interpretation of the dispersionless 
KP hierarchy, we now turn to the dispersionless Toda hierarchy.

\section{Dispersionless Toda hierarchy}

\subsection{Lax formalism} 

The Lax formalism of the dispersionless Toda hierarchy 
\cite{takasaki-takebe-91}, too, is based on a two-dimensional 
phase space with coordinates $(P,q)$. $q$ is a continuum limit 
of the lattice coordinate, whereas $P$ corresponds to the unit-shift 
operator. The Poisson bracket is given by 
\beqn
    \{ F, G \} = 
    P \frac{\rd F}{\rd P} \frac{\rd G}{\rd q}
  - \frac{\rd F}{\rd q} P \frac{\rd G}{\rd P}, 
\eeqn
The corresponding 2-form is $d\log P \wedge dq$. In other words, 
$(\log P, q)$ gives a canonical pair.

The dispersionless Toda hierarchy consists of commuting flows 
of two Lax functions $\cL$ and $\cLbar$ with time variables 
$( t = t_1, t_2, \cdots )$ and 
$ ( \tbar = \tbar_1, \tbar_2, \cdots )$. 
The Lax equations are given by
\beqnarray
    \frac{\rd \cL}{\rd t_n} = \{ B_n , \cL \},  
    &\quad& 
    \frac{\rd \cL}{\rd \tbar_n} = \{ \cBbar_n , \cL \}, 
                                             \nonumber \\
    \frac{\rd \cLbar}{\rd t_n} = \{ B_n , \cLbar \},
    &\quad&
    \frac{\rd \cLbar}{\rd \tbar_n} = \{ \cBbar_n , \cLbar \}, 
\eeqnarray
where Lax functions $\cL$ and $\cLbar$ are two formal Laurent series
of a variable $P$ of the form 
\beqn 
    \cL = P + \sum_{n=0}^{\infty} u_{n+1} (q,t,\tbar)P^{-n}, \quad
    \cLbar = \sum_{n=1}^{\infty} {\bar u}_n q,(t,\tbar) P^n, 
\eeqn
Furthermore, $\cB_n$ and $\cBbar_n$ are given by
\beqn 
    \cB_n = (\cL^n )_{\geq 0}  \quad 
    \cBbar_n = ( \cLbar^{-n})_{\leq -1}, 
\eeqn
where $(\quad)_{\geq 0}$ and $(\quad)_{\leq -1}$ denote the projection 
of Laurent series of $P$ into positive and negative powers respectively. 
They obey the Zakharov-Shabat equations 
\beqnarray
  && 
    \dfrac{\rd \cB_m}{\rd t_n} - \dfrac{\rd \cB_n}{\rd t_m}
    + \{ \cB_m , \cB_n \} = 0, \nonumber \\
  && 
    \dfrac{\rd \cBbar_m}{\rd \tbar_n} - \dfrac{\rd \cBbar_n}{\rd \tbar_m}
     + \{ \cBbar_m,\cBbar_n \} = 0,   \nonumber \\
  && 
    \dfrac{\rd \cB_m}{\rd \tbar_n} - \dfrac{\rd \cBbar_n}{\rd t_m}
    + \{ \cB_m , \cBbar_n \} = 0. 
\eeqnarray

This Lax formalism of the dispersionless Toda hierarchy can 
be extended by adding two counterparts $\cM$ and $\cMbar$ of 
the $\cM$ of the dispersionless KP hierarchy. $\cM$ and 
$\cMbar$ are Laurent series of the form 
\beqnarray 
   \cM &=& \sum_{n = 1}^{\infty} nt_n \cL^n  + q 
          + \sum_{n =1}^{\infty} v_n (q,t,\tbar)\cL^{-n}, 
   \nonumber \\
   \cMbar &=& -\sum_{n = 1}^{\infty} n\tbar_n \cLbar^{-n} + q 
          + \sum_{n = 1}^{\infty} \vbar_n (q,t,\tbar)\cLbar^n
\eeqnarray
and satisfy the Lax equations 
\beqnarray
   \frac{\rd \cM}{\rd t_n} = \{ \cB_n , \cM \},  
   &\quad& 
   \frac{\rd \cM}{\rd \tbar_n} = \{ \cBbar_n , \cM \}, 
   \nonumber \\
   \frac{\rd \cMbar}{\rd t_n} = \{ \cB_n , \cMbar \}, 
   &\quad& 
   \frac{\rd \cMbar}{\rd \tbar_n} = \{ \cBbar_n , \cMbar \}
\eeqnarray
and the canonical Poisson commutation relation
\beqn 
    \{ \cL , \cM \} = \cL,  \quad  
    \{ \cLbar, \cMbar \} = \cLbar.
\eeqn

\subsection{Twistorial structure} 

The above extended Lax formalism, too, can be reformulated 
\`a la Gindikin \cite{gindikin-82}. Let $\omega$ be the 2-form 
\beqn 
    \omega = d\log P \wedge dq 
      + \sum_{n=1}^{\infty} d\cB_n \wedge dt_n 
      + \sum_{n=1}^{\infty} d\cBbar_n \wedge d\tbar_n. 
\eeqn
The Zakharov-Shabat equations then imply that 
\beqn 
    \omega \wedge \omega = 0, 
\eeqn
and the Lax equations and the canonical Poisson commutation 
relations can be rewritten 
\beqn
    d\log\cL \wedge d\cM  = \omega 
  = d\log\cLbar \wedge d\cMbar. 
\eeqn
Thus $\omega$ is a denegerate symplectic form, and $(\cL,\cM)$ 
and $(\cLbar,\cMbar)$ are two different pairs of Darboux coordinates.

It is now straightforward to derive a Riemann-Hilbert problem 
\cite{takasaki-takebe-91}. The two pairs of Darboux coordinates 
are connected by functional relations of the form 
\beqnarray 
    \cL(P,q,t,\tbar) &=& 
      f( \cLbar(P,q,t,\tbar), \cMbar(P,q,t,\tbar), t,\tbar),  
      \nonumber \\
    \cM(P,q,t,\tbar) &=& 
      g( \cLbar(P,q,t,\tbar), \cMbar(P,q,t,\tbar), t,\tbar ).  
\eeqnarray
The transition functions $f = f(P,q)$ and $g = g(P,q)$ are 
required to satisfy the the (twisted) canonical Poisson 
commutation relations 
\beqn
   \{ f(P,q), g(P,q) \} = f(P,q). 
\eeqn
This gives the Riemann-Hilbert problem. 

One can give a twistorial interpretation of this 
Riemann-Hilbert problem in exactly the same way as in the 
case of the dispersionless KP hierarchy.  Note however that 
that, unlike the case of the dispersionless KP hierarchy, 
$(\cL,\cM)$ and $(\cLbar,\cMbar)$ are both dynamical variables. 

It should be also added that a twistorial formulation of 
the dispersionless Toda equation (the lowest member of the 
hierarchy) is presented by Ward \cite{ward-90} in the 
conventional mini-twistor language \cite{hitchin-82,jones-tod-85}.

\subsection{Multi-time Hamiltonian system} 

Because of the presense of two canonical variable pairs 
$(\cL,\cM)$ and $(\cLbar,\cMbar)$, the Hamilton-Jacobi formalism 
of the dispersionless Toda hierarchy is more complicated than 
the case of the dispersionless KP hierarchy. 

Let $(\lambda,\mu)$ and $(\lambdabar,\mubar)$ be two such pairs 
of parameters. One can define two functions 
$P = P(\lambda,\mu,t,\tbar)$ and $q = q(\lambda,\mu,t,\tbar)$ 
implicitly by 
\beqn
    \cL(P,q,t,\tbar) = \lambda, \quad 
    \cM(P,q,t,\tbar) = \mu, 
\eeqn
and similarly, $P = \Pbar(\lambdabar,\mu,t,\tbar)$ and 
$q = \qbar(\lambdabar,\mubar,t,\tbar)$ by 
\beqn
    \cLbar(P,q,t,\tbar) = \lambdabar, \quad 
    \cMbar(P,q,t,\tbar) = \mubar.
\eeqn
We now have the following result: 

\begin{theorem}
Both $\Bigl(P(\lambda,\mu,t,\tbar), q(\lambda,\mu,t,\tbar)\Bigr)$ 
and $\Bigl(\Pbar(\lambdabar,\mu,t,\tbar),
\qbar(\lambdabar,\mubar,t,\tbar)\Bigr)$ 
satisfy the same multi-time Hamiltonian system 
\beqnarray
    \frac{dP}{dt_n} = P \frac{\rd \cB_n}{\rd q}, 
    &\quad&
    \frac{dq}{dt_n} = - P \frac{\rd \cB_n}{\rd P},    
    \nonumber \\
    \frac{dP}{d\tbar_n} = P \frac{\rd \cBbar_n}{\rd q}, 
    &\quad& 
    \frac{dq}{d\tbar_n} = - P \frac{\rd \cBbar_n}{\rd  P}. 
\eeqnarray
\end{theorem}

{\bf Proof}: The proof is almost the same as the case of the 
dispersionless KP hierarchy. The only difference is the Poisson 
bracket.  (It will be more convenient to do calculations in 
the canonical pair $p = \log P$ and $q$ rather than in $P$ 
and $q$.) 
$\Box$
\bigskip

Since this Hamiltonian system lives on a two-dimensional 
phase space, its trajectories should form just a two-dimensional 
family. Accordingly, the two sets of parameters $(\lambda,\mu)$ 
and $(\lambdabar,\mubar)$ should be functionally related. 
Let us write the functional relations as
\beqn
    \lambda = f(\lambdabar,\mubar), \quad 
    \mu = g(\lambdabar,\mubar). 
\eeqn
This implies that the four functions $(\cL,\cM,\cLbar,\cMbar)$ 
satisfy the functional relations 
\beqnarray
  &&
    \cL(P,q,t,\tbar) = 
      f\Bigl(\cLbar(P,q,t,\tbar), \cMbar(P,q,t,\tbar)\Bigr), 
    \nonumber \\
  &&
    \cM(P,q,t,\tbar) = 
      g\Bigl(\cLbar(P,q,t,\tbar), \cMbar(P,q,t,\tbar)\Bigr). 
\eeqnarray
Thus the Riemann-Hilbert problem of the previous subsection 
can be reproduced from the multi-time Hamiltonian system. 

Twistorial interpretation of these results is quite parallel 
to the case of the dispersionless KP hierarchy.

\subsection{Generating functions} 

The two different parametrizations of trajectories of the multi-time 
Hamiltonian system lead to two different canonical transformations 
and multi-time Hamiltonian systems with zero Hamiltonians. The first 
canonical transformation $(P,q) \mapsto (\lambda,\mu)$ is defined by 
a generating function $S(\lambda,q,t,\tbar)$ as 
\beqnarray
  &&
    \frac{\rd S}{\rd \lambda} = \mu, \quad 
    \frac{\rd S}{\rd q} = \log P, 
    \nonumber \\
  &&
    \dfrac{\rd S}{\rd t_n} = \cB_n, \quad 
    \dfrac{\rd S}{\rd \tbar_n} = \cBbar_n. 
\eeqnarray
The transformed Hamiltonian system is given by 
\beqn
    \dfrac{d\lambda}{dt_n} = 0, \quad 
    \dfrac{d\mu}{dt_n} = 0. 
\eeqn
Similarly, the second canonical transformation 
$(P,q) \mapsto (\lambdabar,\mubar)$ is defined by a generating 
function $\Sbar = \Sbar(\lambdabar,q,t,\tbar)$ as 
\beqnarray
  &&
    \frac{\rd \Sbar}{\rd \lambdabar} = \mubar, \quad 
    \frac{\rd \Sbar}{\rd q} = \log P, 
    \nonumber \\
  &&
    \dfrac{\rd \Sbar}{\rd t_n} = \cB_n, \quad 
    \dfrac{\rd \Sbar}{\rd \tbar_n} = \cBbar_n. 
\eeqnarray
The transformed Hamiltonian system is given by 
\beqn
    \dfrac{d\lambdabar}{dt_n} = 0, \quad 
    \dfrac{d\mubar}{dt_n} = 0. 
\eeqn
These Hamiltonian system can be further mapped to a system of 
Gibbons and Kodama type by a simple change of variables 
$(\lambda,\mu) \mapsto (\lambda,\xi)$ and 
$(\lambdabar,\mubar) \mapsto (\lambdabar,\xibar)$ as:  
\beqn
    \xi = \mu - \sum_{n=1}^\infty n t_n \lambda^n, \quad 
    \xibar = \mubar + \sum_{n=1}^\infty n \tbar_n \lambdabar^{-n}. 
\eeqn

The defining equations of the canonical transformations can be 
rewritten 
\beqnarray 
    dS &=& \mu d\lambda + q 
           + \sum_{n=1}^\infty \cB_n dt_n 
           + \sum_{n=1}^\infty \cBbar_n d\tbar_n , 
    \nonumber \\
    d\Sbar &=& \mubar d\lambdabar + q 
           + \sum_{n=1}^\infty \cB_n dt_n 
           + \sum_{n=1}^\infty \cBbar_n d\tbar_n .
\eeqnarray 
Thus $S$ and $\Sbar$ are, in fact, the same as those which 
are already known \cite{takasaki-takebe-91} and connected 
with the linear problem of the Toda lattice hierarchy  
\cite{takasaki-takebe-93,takasaki-takebe-95}.

\section{Concluding Remarks}

The dispersionless KP and Toda hierarchies are also related to 
another geometric structure -- Frobenius manifold. 

Let us recall the notion of Frobenius algebra and Frobenius 
manifold. For more details, we refer to lecture notes of 
Dubrovin \cite{dubrovin-lecture} and Hitchin \cite{hitchin-lecture}. 

A Frobenius algebra is a finite-dimensional commutative and 
associative algebra $V$ with an identity element $e$ and a 
linear form $\theta \in V^*$ for which 
$g(a,b) = \theta(ab)$, $a,b \in V$, is a nondegenerate 
inner product on $V$.  This determines a symmetric form 
$g \in S^2 V^*$. Furthermore, the multiplicative structure 
$V \otimes V \to V$ determines an element of 
$V \otimes V \otimes V^*$. Identifying $V^* \simeq V$ by the 
nondegenerate inner product, one obtains an element 
$c \in V^* \otimes V^* \otimes V^*$. By commutativity 
$c$ becomes totally symmetric, i.e., $c \in S^3 V^*$. 
The three data $(\theta,g,c)$ conversely determines a 
Frobenius algebra. 

Let $M$ be an $n$-manifold with a smoothly varying structure 
of Frobenius algebra in the tangent space at each point. 
This amounts to giving the following threee data: 
\begin{itemize}
\item $\theta \in C^\infty(T^*M)$
\item $g \in C^\infty(S^2 T^*M)$ 
\item $c \in C^\infty(S^3 T^*M)$ 
\end{itemize}
Let $e$ denote the identity element of the point-wise defined 
Frobenius algebra, and $\nabla$ the covariant derivative of 
the Levi-Civita connection determined by the metric $g$. 
$e$ is a vector field on $M$ and called the Euler vector field. 
A manifold $M$ with these data is called a Frobenius manifold 
if the following conditions are satisfied: 
\begin{enumerate}
\item $(M,g)$ is flat, 
\item the Euler vector field $e$ is covariantly constant, 
\item $\nabla c$ is symmetric. 
\end{enumerate}
The flatness of the metric implies the existence of special 
local coordinates (``flat coordinates'') $t^i$ for which the 
coefficents of metric are constant: 
\beqn
    g = \sum \eta_{ij} dt^i dt^j, \quad 
    \eta_{ij} = \mbox{constant}. 
\eeqn
The third condition above, meanwhile, implies the existence of 
another set of distinguished  coordinates (``canonical 
coordinates''). They are local orthogonal coordinates $u^i$ 
for which the metric is written in the following special 
diagonal form with a potential $\phi$: 
\beqn
    g = \sum \mu_i (du^i)^2, \quad 
    \mu_i = \dfrac{\rd \phi}{\rd u^i}. 
\eeqn
(A metric of this form is called, in general, an Egorov metric.) 

A particularly important class of Frobenius manifolds are 
those with homogeneity. This means that all structure functions 
are homogeneous with respect to the Euler vector field $e$. 

Dubrovin \cite{dubrovin-92} pointed out that the moduli spaces 
of marked Riemann surfaces of arbitrary genus and with an 
arbitrary number of marked points (``Hurwitz spaces'') have 
a Frobenius structure (with homogeneity). 
Two special classes of those Frobenius manifolds emerge as 
solutions of the dispersionless KP and Toda hierarchies. 
(Furthermore, these are examples where the Riemann-Hilbert 
problem can be solved rather explicitly.) 

The first example is related to the dispersionless KP hierarchy. 
Let $M$ be the space of complex polynomials
\beqn 
    E(p) = p^{n+1} + a_1 p^{n-1} + \cdots + a_n. 
\eeqn
This is an affine space and its tangent vectors at the point 
is represented by a polynomial $\Edot(p)$ of the form
\beqn 
    \Edot(p) = \Edot_1 p^{n-1} + \cdots + \Edot_n. 
\eeqn
We define a metric on $M$ by
\beqn 
    g(\Edot_1,\Edot_2) 
    = \res_{p = \infty} \frac{\Edot_1 \Edot_2}{E'}dp, 
    = - \sum_i \res_{p = \alpha_i} \frac{\Edot_1 \Edot_2}{E'}dp, 
\eeqn
where $\alpha_i$'s are the roots of $E'(p) = dE(p)/dp$, and 
the summation is over all those roots. $M$ becomes a Frobenius 
manifold with canonical coordinates $u^i = E(\alpha_i)$.

In the context of the dispersionless KP hierarchy, $E(p)$ 
is related to $\cL$ as: 
\beqn
    E(p) = \cL^{n+1}. 
\eeqn
This gives a ``reduction'' of the dispersionless KP hierarchy 
\cite{krichever-92}, which has only a finite number of unknown 
functions ($a_1,\ldots,a_{n-1}$ and $a_n$ in the above notation). Solving 
this relation for $p$ yields a Laurent series of 
the form 
\beqn
    p = \cL + b_1 \cL^{-1} + b_2 \cL^{-2} + \cdots. 
\eeqn
The first $n$ coefficients $b_1,\ldots,b_n$ then give flat 
coordinates.

Similarly, the family of trigonometric polynomials of the form 
\beqn 
    E(p) = e^{np} + a_1 e^{(n-1)p} + \cdots + a_n + a_{n+1}e^{-p} 
\eeqn
gives another example of Frobenius manifold. If we identify 
$P = e^p$, this corresponds to a reduction of the dispersionless 
Toda hierarchy defined by the relations \cite{takasaki-93} 
\beqn
    \cL^n = E(p) = \cLbar^{-1}. 
\eeqn

A very intriguing open problem is to interpret the ``unreduced'' 
dispersionless KP and Toda hierarchies as a kind of infinite 
dimensional Frobenius manifolds.  This will require a drastic 
modification of the notion of Frobenius manifold.  This situation 
is reminiscent of our twistorial interpretation of these hierarchies 
-- the dispersionless hierarchies have no twistor space in the 
ordinary sense, but still retain a virtual counterpart of twistor 
space and twistor lines (encoded in the Riemann-Hilbert problem) 
as well as the notion of twistor functions and twistor lines 
(related to a multi-time Hamiltonian system). 

The same problem may be raised to Krichever's ``universal 
Whitham hierarchy'' \cite{krichever-94}, which includes all 
Hurwitz spaces of arbitrary genus and number of marked points 
as special 
solutions. 

\begin{acknowledge}

\noindent The authors are grateful to Professor Yuji Kodama 
for fruitful discussions. One of the author (P. G.) would 
like to thank Professor Masaki Kashiwara for invitation to RIMS. 

\end{acknowledge}

\end{document}